\documentclass[aps,a4paper,superscriptaddress,preprintnumbers,showpacs,amsmath,amssymb]{revtex4}
\usepackage{graphicx}
\usepackage{dcolumn}
\usepackage{color}
\usepackage{latexsym,amsfonts}
\usepackage{textcomp}
\usepackage{bm}
\usepackage{mathtools,slashed}

\usepackage{ulem}

\begin{document}

 \title{Lorentz Structure of Vector Part\\ of Matrix Elements of
   Transitions $n \longleftrightarrow p$, Caused by Strong Low--Energy
   Interactions and Hypothesis of Conservation of Charged Vector
   Current}
 
 \author{A. N. Ivanov}\email{ivanov@kph.tuwien.ac.at}
 \affiliation{Atominstitut, Technische Universit\"at Wien,
   Stadionallee 2, A-1020 Wien, Austria}

\begin{abstract}
We analyse the Lorentz structure of the matrix elements of the
transitions ``neutron $\longleftrightarrow$ proton'', induced by the
charged hadronic vector current. We show that the term providing
conservation of the charged hadronic vector current in the sense of
the vanishing matrix element of the divergence of the charged hadronic
vector current of the transitions ``neutron $\longleftrightarrow$
proton'' even for different masses of the neutron and proton (see
T. Leitner {\it et al.}, Phys. Rev. C {\bf 73}, 065502 (2006) and
A. M. Ankowski, arXiv:1601.06169 [hep-ph]) has a dynamical origin,
related to the $G$--even first class current contribution. We show
that because of invariance of strong low--energy interactions under
the $G$--parity transformations, the $G$--odd contribution with the
Lorentz structure $q_{\mu}$, where $q_{\mu}$ is a momentum
transferred, does not appear in the matrix elements of the ``neutron
$\longleftrightarrow$ proton' transitions.
\end{abstract}


\date{\today}

\maketitle

\section{Introduction}
\label{sec:introduction}

In the paper by Leitner {\it et al.} \cite{Leitner2006} (see also
\cite{Leitner2006a,Leitner2006b}) the matrix element of the transition
``neutron $\longrightarrow$ proton'' or $n \to p$, induced by the
charged hadronic vector current $V^{(+)}_{\mu}(0)$, has been written
in the following form
\begin{eqnarray}\label{eq:1}
\langle p(k_p,\sigma_p)|V^{(+)}_{\mu}(0)|n(k_n,\sigma_n)\rangle &=&
\bar{u}_p(k_p,\sigma_p)\Big(\Big(\gamma_{\mu} -
\frac{q_{\mu} \slashed{q}}{q^2}\Big)\,F_1(q^2) +
\frac{i\sigma_{\mu\nu}q^{\nu}}{2 m_N}\,F_2(q^2)\Big)
u_n(k_n,\sigma_n)=\nonumber\\ &=&
\bar{u}_p(k_p,\sigma_p)\Big(\gamma^{\nu} \Big(\eta_{\mu\nu}-
\frac{q_{\mu}q_{\nu}}{q^2}\Big)\,F_1(q^2) +
\frac{i\sigma_{\mu\nu}q^{\nu}}{2 m_N}\,F_2(q^2)\Big)\,
u_n(k_n,\sigma_n),
\end{eqnarray}
where $ \bar{u}_p(k_p,\sigma_p)$ and $u_n(k_n,\sigma_n)$ are the Dirac
bispinor wave functions of the free proton and neutron in the final
and initial states of the transition $n \to p$, $m_N = (m_p + m_n)/2$
is a nucleon mass or an averaged nucleon mass, expressed in terms of
the proton $m_p$ and neutron $m_n$ masses, $\eta_{\mu\nu}$ is the
metric tensor of the Minkowski spacetime, and $\gamma_{\mu}$ and
$\sigma_{\mu\nu} = \frac{i}{2}(\gamma_{\mu}\gamma_{\nu} - \gamma_{\nu}
\gamma_{\mu})$ are the Dirac matrices \cite{Itzykson1980}. Then, $q =
k_p - k_n$ is the momentum transferred, and $F_1(q^2)$ and $F_2(q^2)$
are the form factors. The second term in Eq.(\ref{eq:1}) describes the
contribution of the weak magnetism. The right-hand-side (r.h.s.) of
Eq.(\ref{eq:1}) vanishes after multiplication by a momentum
transferred $q^{\mu}$, i.e.
\begin{eqnarray}\label{eq:2}
q^{\mu}\langle p(k_p,\sigma_p)|V^{(+)}_{\mu}(0)|n(k_n,\sigma_n)\rangle = 0,
\end{eqnarray}
even for $m_p \neq m_n$. Such a property of the matrix element of the
transition $n \to p$ testifies conservation of the charged hadronic
vector current $V^{(+)}_{\mu}(x)$, but only in the sense of the
vanishing matrix element $\langle
p(k_p,\sigma_p)|\partial^{\mu}V^{(+)}_{\mu}(x)|n(k_n,\sigma_n)\rangle
= 0$. This, of course, should not contradict the hypothesis of
conservation of the vector current or the CVC hypothesis by Feynman
and Gell-Mann \cite{Feynman1958}. Recently \cite{Ivanov2017c} we have
shown that the term $(- q_{\mu}\slashed{q}/q^2)\,F_1(q^2)$ is the
contribution of the first class current \cite{Weinberg1958,Lee1956}.

This letter is addressed to the analysis of the dynamical nature of
the term with the Lorentz structure $q_{\mu}\slashed{q}$. As has been
proposed in \cite{Ivanov2017c}, the vector part of the matrix element
of the transition $n \to p$, caused by the contributions of the first
class current only, should be taken in the following general form
\begin{eqnarray}\label{eq:3}
\langle p(k_p,\sigma_p)|V^{(+)}_{\mu}(0)|n(k_n,\sigma_n)\rangle =
\bar{u}_p(k_p,\sigma_p)\Big(\gamma_{\mu} \,F_1(q^2) +
\frac{i\sigma_{\mu\nu}q^{\nu}}{2 m_N}\,F_2(q^2) +
\frac{q_{\mu}\slashed{q}}{m^2_N}\,F_4(q^2) \Big) u_n(k_n,\sigma_n).
\end{eqnarray}
Below we show that the appearance of the term with the Lorentz
structure $q_{\mu}\slashed{q}$ is fully caused by strong low--energy
interactions.

The paper is organized as follows. In section \ref{sec:system} we
propose for the analysis of the dynamical nature of the term with the
Lorentz structure $q_{\mu}\slashed{q}$ to use a strongly coupled $\pi
N$--system with the linear pion--nucleon pseudoscalar interaction. We
show that only the total hadronic isovector vector current, being the
sum of the nucleon and mesonic currents, can be locally conserved. In
section \ref{sec:current} we derive the Lorentz structure of the
matrix element of the transition $n \to p$ using the path--integral
technique. In section \ref{sec:conclusion} we discuss the obtained
results. In Appendix A we calculate the cross sections for the
inelastic electron neutrino--neutron scattering and for the
inverse $\beta$--decay. In order to illustrate the influence of the
contributions of the term with the Lorentz structure $(-
q_{\mu}\slashed{q}/q^2)\,F_1(q^2)$ we neglect the contributions of the
weak magnetism and recoil of the outgoing nucleon, and the radiative
corrections. In Fig.\,\ref{fig:fig2} we plot the relative
contributions of the term $(- q_{\mu}\slashed{q}/q^2)\,F_1(q^2)$. We
show that the processes of the inelastic electron
neutrino--neutron scattering and of the inverse $\beta$--decay are
insensitive to the contributions of the term, responsible for the
vanishing of the matrix elements $\langle
p|\partial^{\mu}V^{(+)}_{\mu}(0)|n\rangle = \langle
n|\partial^{\mu}V^{(-)}_{\mu}(0)|p\rangle = 0$ for different masses of
the neutron and proton. In Appendix B we analyse the dynamical nature
of the Lorentz structure of the matrix element $\langle
p|A^{(+)}_{\mu}(0)|n\rangle$ of the transition $n \to p$, caused by
the charged hadronic axial--vector current. We show that the linear
pion--nucleon pseudoscalar interaction, used for the analysis of the
dynamical nature of the Lorentz structure of the charged hadronic
vector part of the transition $n \to p$, allows to reproduce fully the
standard Lorentz structure of the axial--vector part of the hadronic
$n \to p$ transition \cite{Leitner2006}.

\section{Hadronic vector current of strongly coupled pion--nucleon 
system}
\label{sec:system}

As an example of strongly coupled system we consider the $\pi
N$--system with the simplest linear pseudoscalar interaction
\cite{BD1967}. The Lagrangian of such a system is given by
\cite{BD1967}
\begin{eqnarray}\label{eq:4}
\hspace{-0.3in}{\cal L}_{\pi N}(x) =
\bar{N}(x)(i\gamma^{\mu}\partial_{\mu} - m_N)N(x) +
\frac{1}{2}\,\partial_{\mu}\vec{\pi}(x)\cdot\partial_{\mu}\vec{\pi}(x)
- \frac{1}{2}\,m^2_{\pi}\,\vec{\pi}^{\,2}(x) +
g_{\pi}\bar{N}(x)i\gamma^5 \vec{\tau}\cdot \vec{\pi}(x)\,N(x).
\end{eqnarray}
Here $N(x)$ is the nucleon isospin doublet with components $(p(x),
n(x))$, where $p(x)$ and $n(x)$ are the proton and neutron field
operators, $\vec{\pi}(x) = (\pi^+(x), \pi^0(x), \pi^-(x))$ is the pion
field operator, $m_N$ and $m_{\pi}$ are the nucleon and pion masses,
$g_{\pi}$ is the pion--nucleon coupling constant, $\gamma^5$ is the
Dirac matrix \cite{Itzykson1980}, and $\vec{\tau} = (\tau_1,
\tau_2,\tau_3)$ is the Pauli isospin matrix \cite{BD1967}.  

The Lagrangian Eq.(\ref{eq:4}) is invariant under global isospin
transformations \cite{BD1967}. This, according to Feynman and
Gell--Mann \cite{Feynman1958}, leads to the isovector hadronic vector
current of the $\pi N$ system given by
\begin{eqnarray}\label{eq:5}
\hspace{-0.3in}\vec{V}_{\mu}(x) =
\frac{1}{2}\,\bar{N}(x)\vec{\tau}\gamma_{\mu} N(x) + \vec{\pi}(x)
\times \partial_{\mu}\vec{\pi}(x),
\end{eqnarray}
local conservation of which one may check using the equations of
motion.  The Dirac equation for the nucleon and the Klein--Gordon
equation for the pions are given by
\begin{eqnarray}\label{eq:6}
\hspace{-0.3in}(i\gamma^{\mu}\partial_{\mu} - m_N + g_{\pi} i
\gamma^5\,\vec{\tau}\cdot \vec{\pi}(x))N(x) &=& 0,\nonumber\\
\hspace{-0.3in}(\Box + m^2_{\pi})\vec{\pi}(x) -
g_{\pi}\bar{N}(x)i\gamma^5 \vec{\tau}\,N(x) &=& 0.
\end{eqnarray}
Using the Dirac equation for the nucleon Eq.(\ref{eq:6}) one may show
that the nucleon part of the isovector hadronic vector current
Eq.(\ref{eq:5}) is not conserved
\begin{eqnarray}\label{eq:7}
\hspace{-0.3in}\partial^{\mu}\Big(\frac{1}{2}\,\bar{N}(x)\vec{\tau}
\gamma_{\mu} N(x)\Big) =
\frac{1}{2}\,\partial^{\mu}\bar{N}(x)\gamma_{\mu}\vec{\tau}\,N(x) +
\frac{1}{2}\,\bar{N}(x)\vec{\tau}\gamma_{\mu}\partial^{\mu}N(x) = - 
\vec{\pi}(x)\times g_{\pi}\,\bar{N}(x) i \gamma^5\vec{\tau}\,N(x).
\end{eqnarray}
Hence, in the strongly coupled $\pi N$--system a strong
non--conservation of the nucleon part of the isovector hadronic vector
current is caused by strong low--energy interactions but not by
isospin violation. The divergence of the mesonic part of the isovector
hadronic vector current is equal to
\begin{eqnarray}\label{eq:8}
\partial^{\mu}\Big(\vec{\pi}(x) \times \partial_{\mu}\vec{\pi}(x)\Big)
= \vec{\pi}(x)\times g_{\pi}\,\bar{N}(x) i \gamma^5\vec{\tau}\,N(x).
\end{eqnarray}
Summing up the contributions Eqs.(\ref{eq:7}) and (\ref{eq:8}) we get
$\partial^{\mu}\vec{V}_{\mu}(x) = 0$. This means that in the strongly
coupled $\pi N$--system only the total hadronic isovector vector
current, being the sum of the nucleon and mesonic currents, can be
locally conserved.

\section{Dynamical Lorentz structure of the matrix element of the
 $n \to p$ transition, caused by the hadronic vector current
  Eq.(\ref{eq:5})}
\label{sec:current}

The charged hadronic vector current responsible for the hadronic $n
\to p$ transition is equal to \cite{BD1967}
\begin{eqnarray}\label{eq:9}
\hspace{-0.15in}V^{(+)}_{\mu}(x) = \bar{N}(x)
\tau^{(+)}\gamma_{\mu} N(x) + 2\,\varepsilon^{+
  bc}\,\pi^b(x)\partial_{\mu}\pi^c(x) = \bar{p}(x)\gamma_{\mu} n(x) +
\sqrt{2}\,i\,\Big(\pi^0(x) \partial_{\mu}\pi^-(x) - \pi^-(x)
\partial_{\mu}\pi^0(x)\Big),
\end{eqnarray}
where $\tau^{(+)} = (\tau^1 + i\tau^2)/2$,  $\varepsilon^{+ bc} =
(\varepsilon^{1bc} + i\,\varepsilon^{2bc})/2$ and $\varepsilon^{abc}$
is the Levi-Civita isotensor \cite{BD1967}.  Now we may calculate the
matrix element
\begin{eqnarray}\label{eq:10}
\langle {\rm out},
p(\vec{k}_p,\sigma_p)|V^{(+)}_{\mu}(0)|{\rm in}, n(\vec{k}_n,
\sigma_n)\rangle,
\end{eqnarray}
where $\langle {\rm out}, p(\vec{k}_p,\sigma_p)|$ and $|{\rm in},
n(\vec{k}_n, \sigma_n)\rangle$ are the wave functions of the free
proton and neutron in the final (i.e. out--state at $t \to + \infty$)
and initial (i.e. in--state at $t \to - \infty$) states, respectively
\cite{Itzykson1980}. Using the relation $\langle {\rm out},
p(\vec{k}_p,\sigma_p)| = \langle {\rm in},
p(\vec{k}_p,\sigma_p)|{\mathbb S}$, where ${\mathbb S}$ is the
S--matrix, we rewrite the matrix element Eq.(\ref{eq:10}) as follows
\begin{eqnarray}\label{eq:11}
\langle {\rm out}, p(\vec{k}_p,\sigma_p)|V^{(+)}_{\mu}(0)|{\rm in},
n(\vec{k}_n, \sigma_n)\rangle = \langle {\rm in},
p(\vec{k}_p,\sigma_p)|{\mathbb S} V^{(+)}_{\mu}(0)|{\rm in},
  n(\vec{k}_n, \sigma_n)\rangle.
\end{eqnarray}
Since the transition $n \to p$ is fully induced by strong low--energy
interactions, we define the S--matrix only in terms of strong
low--energy interactions. For simplicity we propose to use only $\pi
N$--system, a dynamics of which is determined by the Lagrangian
Eq.(\ref{eq:4}). The corresponding S--matrix is given by
\cite{Itzykson1980}
\begin{eqnarray}\label{eq:12}
{\mathbb S} = {\rm T}e^{\textstyle i\int d^4x\,{\cal L}_{\pi N N}(x)},
\end{eqnarray}
where ${\rm T}$ is a time--ordering operator and ${\cal L}_{\pi N
  N}(x)$ is equal to
\begin{eqnarray}\label{eq:13}
\hspace{-0.3in}{\cal L}_{\pi NN}(x) = g_{\pi}\bar{N}(x)i\gamma^5 \vec{\tau}\cdot \vec{\pi}(x)\,N(x).
\end{eqnarray}
Plugging Eq.(\ref{eq:12}) into Eq.(\ref{eq:11}) we get
\begin{eqnarray}\label{eq:14}
\langle {\rm in}, p(\vec{k}_p,\sigma_p)|{\mathbb S}
V^{(+)}_{\mu}(0)|{\rm in}, n(\vec{k}_n, \sigma_n)\rangle = \langle
{\rm in}, p(\vec{k}_p,\sigma_p)|{\rm T}\Big(e^{\textstyle i\int
  d^4x\,{\cal L}_{\pi NN}(x)}V^{(+)}_{\mu}(0)\Big)|{\rm in},
n(\vec{k}_n, \sigma_n)\rangle.
\end{eqnarray}
The wave functions of the neutron and proton we determine in terms of
the operators of creation (annihilation)
 \begin{eqnarray}\label{eq:15}
|{\rm in}, n(\vec{k}_n, \sigma_n)\rangle &=& a^{\dagger}_{n,\rm
  in}(\vec{k}_n, \sigma_n)|0\rangle,\nonumber\\ \langle {\rm in},
p(\vec{k}_p,\sigma_p)| &=& \langle 0|a_{p,\rm in}(\vec{k}_p,\sigma_p).
\end{eqnarray}
The operators $(a^{\dagger}_{n,\rm in}(\vec{k}_n, \sigma_n), a_{p,\rm
  in}(\vec{k}_p,\sigma_p)$ and $(a_{n,\rm in}(\vec{k}_n, \sigma_n),
a^{\dagger}_{p,\rm in}(\vec{k}_p,\sigma_p)$ obey standard
anticommutation relations \cite{Itzykson1980}
 \begin{eqnarray}\label{eq:16}
{[a_{n,\rm in}(\vec{k}^{\,'}_n, \sigma'_n),a^{\dagger}_{n,\rm
      in}(\vec{k}_n, \sigma_n)]} &=& (2\pi)^3\, 2
E_n\,\delta^{(3)}(\vec{k}^{\,'}_n - \vec{k}_n)\,\delta_{\sigma'_n
  \sigma_n},\nonumber\\ {[a_{n,\rm in}(\vec{k}^{\,'}_n,
    \sigma'_n),a_{n,\rm in}(\vec{k}_n, \sigma_n)]}&=&
{[a^{\dagger}_{n,\rm in}(\vec{k}^{\,'}_n,
    \sigma'_n),a^{\dagger}_{n,\rm in}(\vec{k}_n, \sigma_n)]} =
0,\nonumber\\ {[a_{p,\rm in}(\vec{k}^{\,'}_p,
    \sigma'_p),a^{\dagger}_{p,\rm in}(\vec{k}_p, \sigma_p)]} &=&
(2\pi)^3\, 2 E_p\,\delta^{(3)}(\vec{k}^{\,'}_p -
\vec{k}_p)\,\delta_{\sigma'_p \sigma_p},\nonumber\\ {[a_{p,\rm
      in}(\vec{k}^{\,'}_p, \sigma'_p),a_{p,\rm in}(\vec{k}_p,
    \sigma_p)]}&=& {[a^{\dagger}_{p,\rm in}(\vec{k}^{\,'}_p,
    \sigma'_n),a^{\dagger}_{p,\rm in}(\vec{k}_p, \sigma_p)]} = 0.
\end{eqnarray}
The vacuum wave function we define as follows $|0\rangle =
|0_N\rangle|0_{\pi}\rangle$, where $|0_N\rangle$ and $|0_{\pi}\rangle$
are the vacuum wave functions of a nucleon and mesons,
respectively. Since there are no mesons in the initial and final
states of the transition $n \to p$, the matrix element
Eq.(\ref{eq:14}) we may rewrite as follows
\begin{eqnarray}\label{eq:17}
\langle {\rm in}, p(\vec{k}_p,\sigma_p)|{\mathbb S}
V^{(+)}_{\mu}(0)|{\rm in}, n(\vec{k}_n, \sigma_n)\rangle = {_N}\langle
{\rm in}, p(\vec{k}_p,\sigma_p)|\langle 0_{\pi}|{\rm T}\Big(e^{i\int
  d^4x\,{\cal L}_{\pi NN}(x)}
V^{(+)}_{\mu}(0)\Big)|0_{\pi}\rangle|{\rm in}, n(\vec{k}_n,
\sigma_n)\rangle_N.
\end{eqnarray}
The wave functions ${_N}\langle {\rm in}, p(\vec{k}_p,\sigma_p)|$ and
$|{\rm in}, n(\vec{k}_n, \sigma_n)\rangle_N$ mean that the operators
$(a^{\dagger}_{n,\rm in}(\vec{k}_n, \sigma_n), a_{p,\rm
  in}(\vec{k}_p,\sigma_p)$ act only on the nucleon vacuum wave
function $|0_N\rangle$. The vacuum expectation value $\langle
0_{\pi}|{\rm T}\Big(e^{\textstyle i\int d^4x\,{\cal L}_{\pi NN}(x)}
V^{(+)}_{\mu}(0)\Big)|0_{\pi}\rangle$ we calculate using the
path--integral technique \cite{Bogoliubov1959}. We rewrite the vacuum
expectation value as follows
\begin{eqnarray}\label{eq:18}
\hspace{-0.3in}&&\langle 0_{\pi}|{\rm T}\Big(e^{i\int d^4x\,{\cal
    L}_{\pi NN}(x)} V^{(+)}_{\mu}(0)\Big)|0_{\pi}\rangle = \nonumber\\
\hspace{-0.3in}&&= {\rm T}\Big(\frac{1}{2}\,\bar{N}(0)\tau^{(+)}\gamma_{\mu}
N(0)\int D\vec{\pi}\,e^{\textstyle i\int
  d^4x\,(\frac{1}{2}\,\partial_{\alpha}\vec{\pi}(x)\cdot
  \partial^{\alpha}\vec{\pi}(x) -
  \frac{1}{2}\,m^2_{\pi}\,\vec{\pi}^{\,2} + g_{\pi}\bar{N}(x)i\gamma^5
  \vec{\tau}\cdot \vec{\pi}(x)\,N(x))}\nonumber\\
\hspace{-0.3in}&&+ 2\,\varepsilon^{+bc}\int D\vec{\pi}\,\pi^b(z)
\partial_{\mu}\pi^c(z)\, e^{\textstyle i\int
  d^4x\,(\frac{1}{2}\,\partial_{\alpha}\vec{\pi}(x)\cdot
  \partial^{\alpha}\vec{\pi}(x) -
  \frac{1}{2}\,m^2_{\pi}\,\vec{\pi}^{\,2} + g_{\pi}\bar{N}(x)i\gamma^5
  \vec{\tau}\cdot \vec{\pi}(x)\,N(x))}\Big|_{z = 0}\Big).
\end{eqnarray}
The integrals are Gaussian. The calculation of the first integral
runs as follows. We transcribe it into the form
\begin{eqnarray}\label{eq:19}
\hspace{-0.3in}&&\int D\vec{\pi}\,e^{\textstyle i\int
  d^4x\,(\frac{1}{2}\,\partial_{\alpha}\vec{\pi}(x)\cdot
  \partial^{\alpha}\vec{\pi}(x) -
  \frac{1}{2}\,m^2_{\pi}\,\vec{\pi}^{\,2} + g_{\pi}\bar{N}(x)i\gamma^5
  \vec{\tau}\cdot \vec{\pi}(x)\,N(x))} = \nonumber\\
\hspace{-0.3in}&& = \int D\vec{\pi}\,e^{\textstyle i\int d^4x\,(-
  \frac{1}{2}\,\vec{\pi}\cdot (\Box + m^2_{\pi} - i0)\,\vec{\pi} +
  g_{\pi}\bar{N}(x)i\gamma^5 \vec{\tau}\cdot \vec{\pi}(x)\,N(x))}.
\end{eqnarray}
Then, we make a shift
\begin{eqnarray}\label{eq:20}
\vec{\pi}(x) \to \vec{\pi}(x) + \frac{1}{\Box_x + m^2_{\pi} -
  i0}\,g_{\pi}\bar{N}(x)i\gamma^5 \vec{\tau}\cdot N(x) =
\vec{\pi}(x) + \int d^4y\,\Delta(x - y)\,g_{\pi}\bar{N}(y)i\gamma^5
\vec{\tau} N(y),
\end{eqnarray}
where $\Delta(x - y)$ is the $\pi$--meson propagator
\cite{BD1967}. The result of the integration is
\begin{eqnarray}\label{eq:21}
\hspace{-0.3in}&&\int D\vec{\pi}\,e^{\textstyle i\int
  d^4x\,(\frac{1}{2}\,\partial_{\alpha}\vec{\pi}(x)\cdot
  \partial^{\alpha}\vec{\pi}(x) -
  \frac{1}{2}\,m^2_{\pi}\,\vec{\pi}^{\,2} + g_{\pi}\bar{N}(x)i\gamma^5
  \vec{\tau}\cdot \vec{\pi}(x)\,N(x))} = \nonumber\\
\hspace{-0.3in}&& = e^{\,\textstyle i\,\frac{1}{2}\,g^2_{\pi}\int
  d^4x d^4y\,\bar{N}(x)i\gamma^5 \vec{\tau} N(x) \cdot \Delta(x -
  y)\,\bar{N}(y)i\gamma^5 \vec{\tau} N(y)}.
\end{eqnarray}
For the integration of the pionic part of the charge hadronic vector
current we use the following procedure. We rewrite the path--integral,
given by the second term in the r.h.s. of Eq.(\ref{eq:18}), with an
external source $\vec{J}(x)$ of the $\pi$--meson field:
\begin{eqnarray}\label{eq:22}
\hspace{-0.3in}&&\varepsilon^{abc}\int D\vec{\pi}\,\pi^b(z)
\partial_{\mu}\pi^c(z)\, e^{\textstyle i\int
  d^4x\,(\frac{1}{2}\,\partial_{\alpha}\vec{\pi}(x)\cdot
  \partial^{\alpha}\vec{\pi}(x) -
  \frac{1}{2}\,m^2_{\pi}\,\vec{\pi}^{\,2} + g_{\pi}\bar{N}(x)i\gamma^5
  \vec{\tau}\cdot \vec{\pi}(x)\,N(x))}\Big|_{z = 0} \to \nonumber\\
\hspace{-0.3in}&&\varepsilon^{abc}\int D\vec{\pi}\,\pi^b(z)
\partial_{\mu}\pi^c(z)\, e^{\textstyle i\int
  d^4x\,(\frac{1}{2}\,\partial_{\alpha}\vec{\pi}(x)\cdot
  \partial^{\alpha}\vec{\pi}(x) -
  \frac{1}{2}\,m^2_{\pi}\,\vec{\pi}^{\,2} + g_{\pi}
  \bar{N}(x)i\gamma^5 \vec{\tau}\cdot \vec{\pi}(x)\,N(x) +
  \vec{J}(x)\cdot \vec{\pi}(x))}\Big|_{z = 0}\to\nonumber\\
\hspace{-0.3in}&&\varepsilon^{abc}\int D\vec{\pi}\,\pi^b(z)
\partial_{\mu}\pi^c(z)\, e^{\textstyle i\int d^4x\,(-
  \frac{1}{2}\,\vec{\pi}(x)\cdot (\Box + m^2_{\pi} - i0)\vec{\pi}(x) +
  g_{\pi}\bar{N}(x)i\gamma^5 \vec{\tau}\cdot \vec{\pi}(x)\,N(x) +
  \vec{J}(x)\cdot \vec{\pi}(x))}\Big|_{z = 0}.
\end{eqnarray}
Then, the pionic fields in the integrand we replace by functional
derivatives with respect to the external source:
\begin{eqnarray}\label{eq:23}
\hspace{-0.3in}&&\varepsilon^{abc}\int D\vec{\pi}\,\pi^b(z)
\partial_{\mu}\pi^c(z)\, e^{\textstyle i\int
  d^4x\,(\frac{1}{2}\,\partial_{\alpha}\vec{\pi}(x)\cdot
  \partial^{\alpha}\vec{\pi}(x) -
  \frac{1}{2}\,m^2_{\pi}\,\vec{\pi}^{\,2} + g_{\pi}
  \bar{N}(x)i\gamma^5 \vec{\tau}\cdot \vec{\pi}(x)\,N(x))}\Big|_{z =
  0} \to\nonumber\\
\hspace{-0.3in}&& \to \varepsilon^{abc}(-i)\frac{\delta}{\delta
  J^b(z)}\frac{\partial}{\partial z^{\mu}}(- i)\frac{\delta}{\delta
  J^c(z)}\nonumber\\
\hspace{-0.3in}&&\,\int D\vec{\pi}\, e^{\textstyle i\int d^4x\,(-
  \frac{1}{2}\,\vec{\pi}(x)\cdot (\Box + m^2_{\pi} - i0)\vec{\pi}(x) +
  g_{\pi} \bar{N}(x)i\gamma^5 \vec{\tau}\cdot \vec{\pi}(x)\,N(x) +
  \vec{J}(x)\cdot \vec{\pi}(x))}\Big|_{z = 0,\vec{J} = 0}.
\end{eqnarray}
For the calculation of the integral over $\vec{\pi}$ we make a change
of variables
\begin{eqnarray}\label{eq:24}
\vec{\pi}(x) &\to& \vec{\pi}(x) + \frac{1}{\Box_x + m^2_{\pi} -
  i0}\,\Big(g_{\pi}\bar{N}(x)i\gamma^5 \vec{\tau}\cdot N(x)+
\vec{J}(x)\Big) =\nonumber\\ &=& \vec{\pi}(x) + \int d^4y\,\Delta(x -
y)\,\Big(g_{\pi}\bar{N}(y)i\gamma^5 \vec{\tau} N(y) +
\vec{J}(y)\Big).
\end{eqnarray}
As a result, for the integral over $\vec{\pi}$ we obtain the following
expression
\begin{eqnarray}\label{eq:25}
\hspace{-0.3in}&&\int D\vec{\pi}\, e^{\textstyle i\int d^4x\,(-
  \frac{1}{2}\,\vec{\pi}(x)\cdot (\Box + m^2_{\pi} - i0)\vec{\pi}(x) +
  g_{\pi}\bar{N}(x)i\gamma^5 \vec{\tau}\cdot \vec{\pi}(x)\,N(x) +
  \vec{J}(x)\cdot \vec{\pi}(x))} =\nonumber\\
\hspace{-0.3in}&& = e^{\,\textstyle i\,\frac{1}{2}\,\int d^4x
  d^4y\,(g_{\pi}\bar{N}(x)i\gamma^5 \vec{\tau} N(x) +
  \vec{J}(x))\cdot \Delta(x - y)\,(g_{\pi}
  \bar{N}(y)i\gamma^5 \vec{\tau} N(y) + \vec{J}(y))}.
\end{eqnarray}
Plugging Eq.(\ref{eq:25}) into Eq.(\ref{eq:23}) and calculating the
functional derivatives with respect to external sources we arrive at
the expression
\begin{eqnarray}\label{eq:26}
\hspace{-0.3in}&&\varepsilon^{abc}\int D\vec{\pi}\,\pi^b(z)
\partial_{\mu}\pi^c(z)\, e^{\textstyle i\int
  d^4x\,(\frac{1}{2}\,\partial_{\alpha}\vec{\pi}(x)\cdot
  \partial^{\alpha}\vec{\pi}(x) -
  \frac{1}{2}\,m^2_{\pi}\,\vec{\pi}^{\,2} + g_{\pi
    NN}\bar{N}(x)i\gamma^5 \vec{\tau}\cdot
  \vec{\pi}(x)\,N(x))}\Big|_{z = 0} =\nonumber\\
\hspace{-0.3in}&& = \varepsilon^{abc}\,g^2_{\pi}\int
d^4x\,\Delta(x)\,\bar{N}(x)i\gamma^5 \tau^b\,N(x)\int
d^4y\,\bar{N}(y)i\gamma^5
\tau^c\,N(y)\,\partial_{\mu}\Delta(- y)\nonumber\\
\hspace{-0.3in}&& \times \,e^{\,\textstyle i\,\frac{1}{2}\,g^2_{\pi}\int
  d^4x' d^4y'\,\bar{N}(x')i\gamma^5 \vec{\tau} N(x') \cdot \Delta(x' -
  y')\,\bar{N}(y')i\gamma^5 \vec{\tau} N(y')}.
\end{eqnarray}
Thus, after the calculation of the vacuum expectation value
Eq.(\ref{eq:18}) the matrix element Eq.(\ref{eq:11}) of the transition
$n \to p$ becomes equal to
\begin{eqnarray}\label{eq:27}
\hspace{-0.3in}&&\langle {\rm out},
p(\vec{k}_p,\sigma_p)|V^{(+)}_{\mu}(0)|{\rm in}, n(\vec{k}_n,
\sigma_n)\rangle ={_N}\langle {\rm in}, p(\vec{k}_p,\sigma_p)|{\rm
  T}\Big\{\Big(\bar{N}(0)\tau^{(+)}\gamma_{\mu}N(0) +
2\,\varepsilon^{+bc} g^2_{\pi}\int d^4x \Delta(x) \bar{N}(x) i\gamma^5
\tau^b N(x)\nonumber\\
\hspace{-0.3in}&&\times \int d^4x\,\bar{N}(y)i\gamma^5
\tau^c\,N(y)\,\partial_{\mu}\Delta(- y)\Big)\,e^{\,\textstyle
  i\,\frac{1}{2}\,g^2_{\pi}\int d^4x' d^4y'\,\bar{N}(x')i\gamma^5
  \vec{\tau} N(x') \cdot \Delta(x' - y')\,\bar{N}(y')i\gamma^5
  \vec{\tau} N(y')}\Big\}|{\rm in}, n(\vec{k}_n, \sigma_n)\rangle_N
\nonumber\\
\hspace{-0.3in}&&
\end{eqnarray}
As the first step towards the analysis of the Lorentz structure of the
matrix element of the transition $n \to p$, given by Eq.(\ref{eq:27}),
we propose to consider the contributions of order $g^2_{\pi}$. We
understand that the value of the coupling constant $g_{\pi}$ is
sufficiently large. Nevertheless, the Lorentz structure of the matrix
element Eq.(\ref{eq:27}) can be fully understood to order $g^2_{\pi}$
\cite{Landau1959}.

\subsection{Lorentz structure of the matrix element Eq.(\ref{eq:27})
 to order $g^2_{\pi}$, determined by the mesonic part of the charged
 hadronic vector current Eq.(\ref{eq:9})}

To order $g^2_{\pi}$ the contribution of the mesonic part of the
charged hadronic vector current is given by the expression
\begin{eqnarray}\label{eq:28}
\hspace{-0.3in}&&{_N}\langle {\rm in}, p(\vec{k}_p,\sigma_p)| {\rm
  T}\Big(2\,\varepsilon^{+bc} g^2_{\pi}\int d^4x \Delta(x)
\bar{N}(x)i\gamma^5 \tau^b N(x)\int d^4x\,\bar{N}(y)i\gamma^5
\tau^c\,N(y)\,\partial_{\mu}\Delta(- y)\Big)|{\rm in}, n(\vec{k}_n,
\sigma_n)\rangle_N =\nonumber\\
\hspace{-0.3in}&&= \bar{u}_p(\vec{k}_p, \sigma_p)\,4i g^2_{\pi}\int
d^4x\,\Delta(x)\,e^{\,i k_p\cdot x} i\gamma^5 \int d^4y\,(-i) S_F(x -
y)\,i \gamma^5\,\partial_{\mu} \Delta(- y)\,e^{\,- i k_n\cdot
  y}\,u_n(\vec{k}_n,\sigma_n)\nonumber\\
\hspace{-0.3in}&&- \bar{u}_p(\vec{k}_p, \sigma_p)\,4i g^2_{\pi}\int
d^4x\,\partial_{\mu}\Delta(- x)\,e^{\,i k_p\cdot x} i\gamma^5 \int
d^4y\,(-i) S_F(x - y)\,i \gamma^5\,\Delta(y)\,e^{\,-
  i k_n\cdot y}\,u_n(\vec{k}_n,\sigma_n),
\end{eqnarray}
where $S_F(x - y)$ is the nucleon propagator \cite{BD1967}. For the
derivation of Eq.(\ref{eq:28}) we have used the relation
$\varepsilon^{+bc}\tau^b\tau^c = 2 i \tau^{(+)}$.  In the momentum
representation the r.h.s. of Eq.(\ref{eq:28}) reads
\begin{eqnarray}\label{eq:29}
\hspace{-0.3in}&&{_N}\langle {\rm in}, p(\vec{k}_p,\sigma_p)| {\rm
  T}\Big(2\,\varepsilon^{+bc} g^2_{\pi}\int d^4x \Delta(x)
\bar{N}(x)i\gamma^5 \tau^b N(x)\int d^4x\,\bar{N}(y)i\gamma^5
\tau^c\,N(y)\,\partial_{\mu}\Delta(- y)\Big)|{\rm in}, n(\vec{k}_n,
\sigma_n)\rangle_N =\nonumber\\
\hspace{-0.3in}&&= \bar{u}_p(\vec{k}_p,
\sigma_p)\,\frac{g^2_{\pi}}{4\pi^2}\int \frac{d^4p}{\pi^2
  i}\,\frac{1}{m^2_{\pi} - (p - k_p)^2 - i0}\,\gamma^5 \frac{1}{m_N -
  \hat{p} - i0}\,\gamma^5 \frac{(k_p + k_n - 2 p)_{\mu}}{m^2_{\pi} -
  (p - k_n)^2 - i0}\,u_n(\vec{k}_n,\sigma_n).
\end{eqnarray}
The integral is symmetric with respect to the transformation $k_p
\longleftrightarrow k_n$. This means that the momentum integral
possesses the following Lorentz structure
\begin{eqnarray}\label{eq:30}
\hspace{-0.3in} \int \frac{d^4p}{\pi^2 i} \,\frac{1}{m^2_{\pi} - (p -
  k_p)^2 - i0}\,\gamma^5 \frac{1}{m_N - \hat{p} - i0}\,\gamma^5
\frac{(k_p + k_n - 2 p)_{\mu}}{m^2_{\pi} - (p - k_n)^2 - i0} =
a\,\gamma_{\mu} + b\,P_{\mu}\slashed{P} + c\,q_{\mu} \slashed{q},
\end{eqnarray}
which can be confirmed by a direct calculation of the integral, where
$P = (k_p + k_n)/2$ and $a$, $b$ and $c$ are coefficients, which can
be determined by a direct calculation of the integral. The symmetry of
the integral with respect to the transformation $k_p
\longleftrightarrow k_n$ testifies that the term with the Lorentz
structure $q_{\mu} = (k_p - k_n)_{\mu}$, which is antisymmetric with
respect to the transformation $k_p \longleftrightarrow k_n$, does not
appear in the matrix element of the transition $n \to p$ in agreement
with a suppression of the contributions of the second class currents
\cite{Weinberg1958,Lee1956}. 

As a consequence of the relations $m^2_N \gg m^2_{\pi} \gg q^2$ one
may perform the calculation of the momentum integral Eq.(\ref{eq:30})
in the heavy baryon approximation \cite{Ericson2006}.  Since we are
interested in the term with the Lorentz structure $q_{\mu} \slashed{q}$
only, skipping standard intermediate steps of the calculations for the
coefficient $c$ we obtain the following result
\begin{eqnarray}\label{eq:31}
\hspace{-0.3in}c = \frac{1}{6m_N
  m_{\pi}}\,\arctan\Big(\frac{m_N}{m_{\pi}} \Big),
\end{eqnarray}
where we have neglected the contributions of order $O(1/m^2_N)$. Then,
using the Dirac equations $\bar{u}_p\slashed{P} u_n = (m_p + m_n)/2 = m_N$
that does not violate the property of the term with the Lorentz
structure $P_{\mu} \slashed{P}$ to be a contribution of the first class
current \cite{Ivanov2017c} and the Gordon identity \cite{Itzykson1980}
\begin{eqnarray}\label{eq:32}
\bar{u}_p(\vec{k}_p, \sigma_p)\,\frac{(k_p +
  k_p)_{\mu}}{2m_N}\,u_n(\vec{k}_n,\sigma_n) = \bar{u}_p(\vec{k}_p,
\sigma_p)\,\gamma_{\mu}\,u_n(\vec{k}_n,\sigma_n) -
\bar{u}_p(\vec{k}_p,
\sigma_p)\,\frac{i\sigma_{\mu\nu}q^{\nu}}{2m_N}\,u_n(\vec{k}_n,\sigma_n)
\end{eqnarray}
we transcribe Eq.(\ref{eq:29}) into the form
\begin{eqnarray}\label{eq:33}
\hspace{-0.3in}&&{_N}\langle {\rm in}, p(\vec{k}_p,\sigma_p)| {\rm
  T}\Big( 2\,\varepsilon^{+bc} g^2_{\pi}\int d^4x \Delta(x)
\bar{N}(x)i\gamma^5 \tau^b N(x)\int d^4x\,\bar{N}(y)i\gamma^5
\tau^c\,N(y)\,\partial_{\mu}\Delta(- y)\Big)|{\rm in}, n(\vec{k}_n,
\sigma_n)\rangle_N =\nonumber\\
\hspace{-0.3in}&&= \bar{u}_p(\vec{k}_p,
\sigma_p)\,\frac{g^2_{\pi}}{4\pi^2}\Big(A_{\pi}\,\gamma_{\mu} +
  B_{\pi}\,\frac{i\sigma_{\mu\nu}q^{\nu}}{2 m_N} +
  C_{\pi}\,\frac{q_{\mu}\slashed{q}}{m^2_N}\Big)\,u_n(\vec{k}_n,\sigma_n),
\end{eqnarray}
where $A_{\pi} = a + m_N b$, $B_{\pi} = - m_N b$ and $C_{\pi} = m^2_N
c$.

\subsection{Lorentz structure of the matrix element Eq.(\ref{eq:27}) 
to order $g^2_{\pi}$, determined by the nucleon part of the charged
hadronic vector current Eq.(\ref{eq:9})}

To order $g^2_{\pi}$ the dynamical contribution of the nucleon part of
the charge hadronic vector current to the matrix element of the
transition $n \to p$, given by Eq.(\ref{eq:27}), is determined by the
matrix element
\begin{eqnarray}\label{eq:34}
\hspace{-0.3in}&&{_N}\langle {\rm in}, p(\vec{k}_p,\sigma_p)|{\rm
  T}\Big(\bar{N}(0)\tau^{(+)}\gamma_{\mu}N(0)
i\,\frac{1}{2}\,g^2_{\pi}\int d^4x d^4y\,\bar{N}(x)i\gamma^5
\vec{\tau} N(x) \cdot \Delta(x - y)\,\bar{N}(y)i\gamma^5 \vec{\tau}
N(y)\Big)|{\rm in}, n(\vec{k}_n, \sigma_n)\rangle_N =\nonumber\\
\hspace{-0.3in}&& = \bar{u}_p(\vec{k}_p, \sigma_p) 3 i g^2_{\pi}\int
d^4x d^4y\,\gamma_{\mu}(-i)S_F(-x) i\gamma^5 (-i) S_F(x - y) \Delta(x - y)
i\gamma^5 e^{\,-i k_n\cdot y}\,u_n(\vec{k}_n, \sigma_n)\nonumber\\
\hspace{-0.3in}&& + \bar{u}_p(\vec{k}_p, \sigma_p) 3 i g^2_{\pi}\int
d^4x d^4y\,e^{\,i k_p\cdot x}i\gamma^5 (-i) S_F(x - y) \Delta(x -
y)i\gamma^5 (-i) S_F(y)\,\gamma_{\mu} \,u_n(\vec{k}_n, \sigma_n)\nonumber\\
\hspace{-0.3in}&& - \bar{u}_p(\vec{k}_p, \sigma_p) i g^2_{\pi}\int
d^4x d^4y\,e^{\,i k_p\cdot x}i\gamma^5 (-i) S_F(x)
\gamma_{\mu}\Delta(x - y) (-i) S_F(-y) i\gamma^5\,e^{\,-i k_n\cdot
  y}\,u_n(\vec{k}_n, \sigma_n),
\end{eqnarray}
where we have used the relations $\vec{\tau}^{\,2} = 3$ and
$\vec{\tau}\cdot \tau^{(+)}\vec{\tau} = - \tau^{(+)}$. The
contributions of the first two terms in Eq.(\ref{eq:34}) can be
removed by renormalization of the masses and wave functions of the
neutron and proton, respectively
\cite{Matthews1951,Schweber1962}. Thus, a non--trivial contribution
comes from the third term only. In the momentum representation it
reads
\begin{eqnarray}\label{eq:35}
\hspace{-0.3in}&&{_N}\langle {\rm in}, p(\vec{k}_p,\sigma_p)|{\rm
  T}\Big(\bar{N}(0)\tau^{(+)}\gamma_{\mu}N(0)
i\,\frac{1}{2}\,g^2_{\pi}\int d^4x d^4y\,\bar{N}(x)i\gamma^5
\vec{\tau} N(x) \cdot \Delta(x - y)\,\bar{N}(y)i\gamma^5 \vec{\tau}
N(y)\Big)|{\rm in}, n(\vec{k}_n, \sigma_n)\rangle_N =\nonumber\\
\hspace{-0.3in}&& = \bar{u}_p(\vec{k}_p, \sigma_p)
\frac{g^2_{\pi}}{16\pi^2}\int\frac{d^4p}{\pi^2i}\,\gamma^5\,\frac{1}{m_N
  - \hat{k}_p + \hat{p} - i0}\,\gamma_{\mu}\,\frac{1}{m_N - \hat{k}_n
  + \hat{p} - i0}\gamma^5\,\frac{1}{m^2_{\pi} - p^2 - i0}
\,u_n(\vec{k}_n, \sigma_n).
\end{eqnarray}
This integral is also symmetric with respect to the transformation
$k_p \longleftrightarrow k_n$, so it should also have a structure
\begin{eqnarray}\label{eq:36}
\hspace{-0.3in}\int\frac{d^4p}{\pi^2i}\,\gamma^5\,\frac{1}{m_N -
  \hat{k}_p + \hat{p} - i0}\,\gamma_{\mu}\,\frac{1}{m_N - \hat{k}_n +
  \hat{p} - i0}\gamma^5\,\frac{1}{m^2_{\pi} - p^2 - i0} =
a'\,\gamma_{\mu} + b'\,\slashed{P} P_{\mu} + c'\,q_{\mu} \slashed{q},
\end{eqnarray}
where the coefficients $a'$, $b'$ and $c'$ are determined by a direct
calculation of the integral. Thus, the contribution of the term with
the Lorentz structure $q_{\mu} = (k_p - k_n)_{\mu}$, which is
antisymmetric with respect to the transformation $k_p
\longleftrightarrow k_n$, does not appear in the nucleon part of the
charge hadronic vector current.  A direct calculation of the integral
in Eq.(\ref{eq:36}) gives the following value of the coefficient $c'$:
$c' = - 1/24 m^2_N$. Using the Dirac equations $\bar{u}_p \slashed{P}
u_n = m_N$ that does not violate the property of the term with the
Lorentz structure $P_{\mu} \slashed{P}$ to be a contribution of the
first class current \cite{Ivanov2017c} and the Gordon identity
Eq.(\ref{eq:32}) we transcribe the r.h.s. of Eq.(\ref{eq:35}) into the
form
\begin{eqnarray}\label{eq:37}
\hspace{-0.3in}&&{_N}\langle {\rm in}, p(\vec{k}_p,\sigma_p)|{\rm
  T}\Big(\bar{N}(0)\tau^{(+)}\gamma_{\mu}N(0)
i\,\frac{1}{2}\,g^2_{\pi}\int d^4x d^4y\,\bar{N}(x)i\gamma^5
\vec{\tau} N(x) \cdot \Delta(x - y)\,\bar{N}(y)i\gamma^5 \vec{\tau}
N(y)\Big)|{\rm in}, n(\vec{k}_n, \sigma_n)\rangle_N =\nonumber\\
\hspace{-0.3in}&& = \bar{u}_p(\vec{k}_p, \sigma_p)\,
\frac{g^2_{\pi}}{4\pi^2}\,\Big(A_N\,\gamma_{\mu} +
B_N\,\frac{i\sigma_{\mu\nu}q^{\nu}}{2m_N} +
C_N\,\frac{q_{\mu}\slashed{q}}{m^2_N}\Big)\, u_n(\vec{k}_n, \sigma_n).
\end{eqnarray}
where $A_N = (a' + m_Nb')/4$, $B_N = - m_N b'/4$ and $C_N = m^2_N c'$.

Summing up the contributions of the nucleon and mesonic parts of the
charged hadronic vector current for the vector part of the matrix
element of the transition $n \to p$ Eq.(\ref{eq:27}), calculated to
order $g^2_{\pi}$, we obtain the expression
\begin{eqnarray}\label{eq:38}
\hspace{-0.3in}&&\langle {\rm out},
p(\vec{k}_p,\sigma_p)|V^{(+)}_{\mu}(0)|{\rm in}, n(\vec{k}_n,
\sigma_n)\rangle = {_N}\langle {\rm in}, p(\vec{k}_p,\sigma_p)|{\rm
  T}\Big\{\Big(\bar{N}(0)\tau^{(+)}\gamma_{\mu}N(0) +
2\,\varepsilon^{+bc} g^2_{\pi}\int d^4x \Delta(x) \bar{N}(x) i\gamma^5
\tau^b N(x)\nonumber\\
\hspace{-0.3in}&&\times \int d^4x\,\bar{N}(y)i\gamma^5
\tau^c\,N(y)\,\partial_{\mu}\Delta(- y)\Big)\,e^{\,\textstyle
  i\,\frac{1}{2}\,g^2_{\pi}\int d^4x' d^4y'\,\bar{N}(x')i\gamma^5
  \vec{\tau} N(x') \cdot \Delta(x' - y')\,\bar{N}(y')i\gamma^5
  \vec{\tau} N(y')}\Big\}|{\rm in}, n(\vec{k}_n, \sigma_n)\rangle_N
\nonumber\\
\hspace{-0.3in}&&= \bar{u}_p(\vec{k}_p, \sigma_p)\Big(\Big(1 +
\frac{g^2_{\pi}}{4\pi^2}(A_N + A_{\pi})\Big)\,\gamma_{\mu} + \frac{
  g^2_{\pi}}{4\pi^2}\,(B_N + B_{\pi})\,\frac{i\sigma_{\mu\nu}
  q^{\nu}}{2 m_N} + \frac{g^2_{\pi}}{4\pi^2}\,(C_N + C_{\pi})\,
\frac{q_{\mu}\slashed{q}}{M^2_{\pi}}\Big)\, u_n(\vec{k}_n, \sigma_n).
\end{eqnarray}
Thus, we have shown that the matrix element of the transition $n \to
p$, calculated to order $g^2_{\pi}$, can be expressed in terms of
three Lorentz structures $\gamma_{\mu}$, $i\sigma_{\mu\nu}q^{\nu}$ and
$q_{\mu}\slashed{q}$, which are induced by the first class current
\cite{Ivanov2017c}. Indeed, the isovector hadronic vector current
Eq.(\ref{eq:5}) has a positive $G$--parity and belongs to the first
class current \cite{Ivanov2017c}
\begin{eqnarray}\label{eq:39}
\hspace{-0.3in}&&\vec{V}_{\mu}(x) \stackrel{G}{\longrightarrow} G
\vec{V}_{\mu}(x) G^{-1} = G\Big(\bar{N}(x)
\frac{1}{2}\,\vec{\tau}\,\gamma_{\mu} N(x) \Big)G^{-1} +
G\Big(\vec{\pi}(x)\times \partial_{\mu}\vec{\pi}(x)\Big)G^{-1} =\nonumber\\
\hspace{-0.3in}&&= N^T(x)C (- i)\tau_2 \vec{\tau}\,\gamma_{\mu}
i\tau_2 C \bar{N}^T(x) + G\vec{\pi}(x) G^{-1} \times G\partial_{\mu}
\vec{\pi}(x) G^{-1} = \vec{V}_{\mu}(x),
\end{eqnarray}
where $T$ is a transposition and $C$ is the charge conjugation matrix
\cite{Itzykson1980}. For the derivation of the relation
Eq.(\ref{eq:39}) we have used the relations $C\gamma_{\mu}C =
\gamma^T_{\mu}$, $i\tau_2 \vec{\tau}i\tau_2 = \vec{\tau}^{\,T}$ and
$N^T(x) \bar{N}^T(x) = - \bar{N}(x) N(x)$ \cite{Itzykson1980} and
$G\vec{\pi}(x) G^{-1} = - \vec{\pi}(x)$
\cite{Weinberg1958,Lee1956}. 

Since, the coefficient $C_N$ is much smaller than the coefficient
$C_{\pi}$, the contribution of the Lorentz structure $q_{\mu}
\slashed{q}$ to the matrix element of the transition $n \to p$ is
practically determined by the mesonic part of the charged hadronic
vector current Eq.(\ref{eq:9}). Of course, the coefficients $A_{\pi}$
and $A_N$ can depend on the ultra--violet cut--off $\Lambda$. However,
such a dependence can be removed by renormalization of the coupling
constant $g^2_{\pi}$ \cite{Matthews1951,Schweber1962}.

As strong low--energy interactions are invariant under the $G$--parity
transformation \cite{Lee1956} (see also \cite{Ivanov2017c})
\begin{eqnarray}\label{eq:40}
\hspace{-0.3in}&&{\cal L}_{\pi NN}(x) = g_{\pi} \bar{N}(x)i\gamma^5
\vec{\tau} N(x)\cdot \vec{\pi}(x) \stackrel{G}{\longrightarrow}
G\Big(g_{\pi} \bar{N}(x)i\gamma^5 \vec{\tau}N(x)\cdot
\vec{\pi}(x)\Big)G^{-1} = \overline{N^G}(x) i\gamma^5 \vec{\tau}
N^G(x)\cdot \vec{\pi}^{\,G}(x) =\nonumber\\
\hspace{-0.3in}&&= N^T(x) C (-i)\tau_2 i \gamma^5 \vec{\tau} i\tau_2 C
\bar{N}^T(x) \cdot \Big(- \vec{\pi}(x)\Big) = {\cal L}_{\pi NN}(x),
\end{eqnarray}
where we have used the relation $C \gamma^5 C = - \gamma^{5 T}$
\cite{Itzykson1980}, and the terms with the Lorentz structures
$\gamma_{\mu}$, $i\sigma_{\mu\nu}q^{\nu}$ and $q_{\mu}\slashed{q}$
possess the positive $G$--parity, i.e. they are the contributions of
the first class current \cite{Ivanov2017c}, the term with the Lorentz
structure $q_{\mu}$, having a negative $G$--parity \cite{Ivanov2017c},
should not appear in the matrix element of the transition $n \to p$
Eq.(\ref{eq:27}) to any order of $g^2_{\pi}$--expansion. This allows
to write \cite{Ivanov2017c}
\begin{eqnarray}\label{eq:41}
\langle {\rm out}, p(\vec{k}_p,\sigma_p)|V^{(+)}_{\mu}(0)|{\rm in},
n(\vec{k}_n, \sigma_n)\rangle = \bar{u}_p(\vec{k}_p,
\sigma_p)\,\Big(F_1(q^2)\,\gamma_{\mu} +
F_2(q^2)\,\frac{i\sigma_{\mu\nu}q^{\nu}}{2m_N} +
F_4(q^2)\,\frac{q_{\mu}\slashed{q}}{m^2_N}\Big)\,u_n(\vec{k}_n,\sigma_n),
\end{eqnarray}
where $F_1(q^2)$, $F_2(q^2)$ and $F_4(q^2)$ are form factors,
calculated to all orders of $g^2_{\pi}$--expansion.

\section{Conclusive discussions}
\label{sec:conclusion}

We have analysed the Lorentz structure of the matrix element of the
transition $n \to p$, caused by the charged hadronic vector
current. We have shown that in addition to the standard terms with the
Lorentz structure $F_1(q^2)\,\gamma_{\mu}$ and
$F_2(q^2)\,i\sigma_{\mu\nu}q^{\nu}/2 m_N$, caused by the contributions
of the electric charge distribution and the weak magnetism inside the
hadron, one obtains the term with the structure
$F_4(q^2)\,q_{\mu}\slashed{q}/m^2_N$. Using the simplest model of
strongly coupled $\pi N$--system with the linear pion--nucleon
pseudoscalar interaction we have shown that the contribution of the
term with the Lorentz structure $F_4(q^2)\,q_{\mu}\slashed{q}/M^2$ is
practically induced by the mesonic part of the hadronic isovector
vector current. We have also shown that the term with the Lorentz
structure $F_3(q^2)\,q_{\mu}/m_N$, caused by the second class current
\cite{Weinberg1958,Lee1956}, cannot be, in principle, induced by strong
low--energy interactions invariant under $G$--parity transformations.

A requirement of conservation of the charged hadronic vector current
even for different masses of the hadrons in the initial and final
states (see \cite{Leitner2006, Leitner2006a, Leitner2006b,
  Paschos2006a, Paschos2006b} and so on) in the sense of the vanishing
of the matrix element $\langle
p|\partial^{\mu}V^{(+)}_{\mu}(0)|n\rangle = 0$ of the hadronic $n \to
p$ transition leads to the relation $F_4(q^2) = -
(m^2_N/q^2)\,F_1(q^2)$.  Such a relation leads to the appearance of
the term $(- q_{\mu}\slashed{q}/q^2)\,F_1(q^2)$ in the matrix element
of the hadronic $n \to p$ transition.

For simplicity we have restricted our analysis by the simplest theory
of $\pi N$ strong interactions described by the Lagrangian
Eq.(\ref{eq:4}) with the linear pseudoscalar $\pi NN$--interaction
\cite{BD1967,Matthews1951,Schweber1962}. However, one may assert that
the obtained result, i.e. the existence of the term with the Lorentz
structure $q_{\mu} \slashed{q}$ and the suppression of the term with
the Lorentz structure $q_{\mu}$, which are the contributions of the
first and second class currents, respectively, should be valid in any
theory of strong low--energy interactions
\cite{Weinberg1968,Weinberg1968a,Scherer2012}, which are invariant
under $G$--transformations \cite{Lee1956} (see also
\cite{Weinberg1958}). Our assertion is based only on $G$--invariance
of such theories. Indeed, it is hardly possible to perform analytical
calculations, which are similar to those we have carried out in this
paper, within such complicated non--linear theories
of meson--nucleon low--energy interactions as
\cite{Weinberg1968,Weinberg1968a} and Chiral perturbation theory
\cite{Scherer2012}.

In Appendix A we have shown that the cross sections for the electron
neutrino--neutron scattering and for the inverse $\beta$--decay,
calculated in the non--relativistic approximation with respect to the
outgoing hadron, are insensitive to the contributions of the term $(-
q_{\mu}\slashed{q}/q^2)\,F_1(q^2)$. That is why one may assert that it
is important to search for processes, which are sensitive to the
contributions of the term $(- q_{\mu}\slashed{q}/q^2)\,F_1(q^2)$.

In Appendix B we have analysed the
dynamical nature of the Lorentz structure of the matrix element
$\langle p|A^{(+)}_{\mu}(0)|n\rangle$ of the transition $n \to p$,
caused by the charged hadronic axial--vector current
$A^{(+)}_{\mu}(0)$. We have shown that the low--energy pion--nucleon
interaction Eq.(\ref{eq:13}) allows to reproduce fully the standard
Lorentz structure of the axial--part of the hadronic $n \to p$
transition \cite{Leitner2006}.

Of course, our results, obtained for the hadronic $n \to p$ transition
\cite{Leitner2006}, are fully valid for the hadronic $p \to n$
transition \cite{Ankowski2016}.

\section{Acknowledgements}
\label{sec:acknowledgements}

This work was supported by the Austrian ``Fonds zur F\"orderung der
Wissenschaftlichen Forschung'' (FWF) under Contracts I689-N16,
P26781-N20 and P26636-N20 and ``Deutsche F\"orderungsgemeinschaft''
(DFG) AB 128/5-2.

\section{Appendix A: Cross sections for the inelastic electron 
neutrino--neutron scattering and for the inverse $\beta$--decay}
\renewcommand{\theequation}{A-\arabic{equation}}
\setcounter{equation}{0}

In this Appendix we calculate the cross sections for the
inelastic scattering $\nu_e + n \to p + e^-$ and the inverse
$\beta$--decay $\bar{\nu}_e + p \to n + e^+$ by taking into account
the contributions of the term $- q_{\mu}\slashed{q}/q^2$ responsible
for the constraint $\langle
h'|\partial^{\mu}V^{(\pm)}_{\mu}(0)|h\rangle = 0$ even for different
masses of incoming $h$ and outgoing $h'$ hadrons. Below the
contributions of such a term we call the contributions of Exact
Conservation of the charged weak hadronic Vector Current or the ECVC
effect.

 The amplitudes of the inelastic scattering $\nu_e + n \to p +
 e^-$ and the inverse $\beta$--decay $\bar{\nu}_e + p \to n + e^+$ we
 define in the non--relativistic approximation for the outgoing
 nucleon. They are equal to
\begin{eqnarray}\label{eq:A.1}
\hspace{-0.3in} M(\nu_e n \to p\,e^-) = -
\frac{G_F}{\sqrt{2}}\,V_{ud}\,\langle p(\vec{k}_p,
\sigma_p)|J^{(+)}_{\mu}(0)|n(\vec{k}_n,
\sigma_n)\rangle\,\Big[\bar{u}_-(\vec{k}_-, \sigma_-)\gamma^{\mu}(1 -
  \gamma^5)\, u_{\nu}(\vec{k}, - \frac{1}{2})\Big]
\end{eqnarray}
and
\begin{eqnarray}\label{eq:A.2}
\hspace{-0.3in} M(\bar{\nu}_e p \to n\,e^+) = +
\frac{G_F}{\sqrt{2}}\,V_{ud}\,\langle n(\vec{k}_n,
\sigma_n)|J^{(-)}_{\mu}(0)|p(\vec{k}_p,
\sigma_p)\rangle\,\Big[\bar{v}_{\bar{\nu}}(\vec{k}_{\bar{\nu}}, +
  \frac{1}{2})\gamma^{\mu}(1 - \gamma^5)\, v_+(\vec{k}_+,
  \sigma_+)\Big],
\end{eqnarray}
where $G_F$ and $V_{ud}$ are the Fermi weak coupling constant and the
Cabibbo--Kobayashi--Maskawa (CKM) matrix element \cite{PDG2016},
$\bar{u}_-(\vec{k}_-,\sigma_-)$ and $u_{\nu}(\vec{k}_{\nu}, -
\frac{1}{2})$ are the Dirac wave functions of the free electron and
electron neutrino with 3--momenta $\vec{k}_-$ and $\vec{k}_{\nu}$ and
polarizations $\sigma_e = \pm 1$ and $- \frac{1}{2}$
\cite{Ivanov2013,Ivanov2014,Ivanov2013a}, respectively, and
$(\gamma_{\mu}, \gamma^5)$ are the Dirac matrices. Then,
$\bar{v}_{\bar{\nu}}(\vec{k}_{\bar{\nu}}, + \frac{1}{2})$ and
$v_+(\vec{k}_+, \sigma_+)$ are the Dirac wave functions of the
electron antineutrino and positron with 3--momenta
$\vec{k}_{\bar{\nu}}$ and $\vec{k}_+$ and polarizations $\sigma_+ =
\pm 1$ and $+ \frac{1}{2}$ \cite{Ivanov2013,Ivanov2014,Ivanov2013a},
respectively. The matrix elements $\langle p(\vec{k}_p,
\sigma_p)|J^{(+)}_{\mu}(0)|n(\vec{k}_n, \sigma_n)\rangle$ and $\langle
n(\vec{k}_n, \sigma_n)|J^{(-)}_{\mu}(0)|p(\vec{k}_p, \sigma_p)\rangle$
of the hadronic $n \to p$ and $p \to n$ transitions we define as
follows \cite{Leitner2006}
\begin{eqnarray}\label{eq:A.3}
\hspace{-0.3in}\langle p(\vec{k}_p, \sigma_p)|J^{(+)}_{\mu}(0)
|n(\vec{k}_n, \sigma_n)\rangle = \bar{u}_p(\vec{k}_p,
\sigma_p)\Big(F_1(q^2)\Big(\gamma_{\mu} -
\frac{q_{\mu}\slashed{q}}{q^2}\Big) +
F_A(q^2)\,\gamma_{\mu}\gamma^5\Big)\,u_n(\vec{k}_n, \sigma_n)
\end{eqnarray}
and 
\begin{eqnarray}\label{eq:A.4}
\hspace{-0.3in}\langle n(\vec{k}_n, \sigma_n)|J^{(-)}_{\mu}(0)
|p(\vec{k}_p, \sigma_p)\rangle = \bar{u}_n(\vec{k}_n,
\sigma_n)\Big(F_1(q^2)\Big(\gamma_{\mu} -
\frac{q_{\mu}\slashed{q}}{q^2}\Big) +
F_A(q^2)\,\gamma_{\mu}\gamma^5\Big)\,u_p(\vec{k}_p, \sigma_p),
\end{eqnarray}
where $J^{(\pm)}_{\mu}(0) = V^{(\pm)}_{\mu}(0) - A^{(\pm)}_{\mu}(0)$,
$u_n(\vec{k}_n, \sigma_n)$ and $u_p(\vec{k}_p, \sigma_p)$ are the
Dirac wave functions of the free neutron and proton with 3--momenta
and polarizations $(\vec{k}_n, \sigma_n = \pm 1)$ and $(\vec{k}_p,
\sigma_p = \pm 1)$. Then, $F_1(q^2)$ and $F_A(q^2)$ are the vector and
axial--vector form factors \cite{Leitner2006}. The vector parts of the
matrix elements Eqs.(\ref{eq:A.3}) and (\ref{eq:A.4}) obey the
constraints
\begin{eqnarray}\label{eq:A.5}
\hspace{-0.3in}\langle p(\vec{k}_p, \sigma_p)|\partial^{\mu}
V^{(+)}_{\mu}(0) |n(\vec{k}_n, \sigma_n)\rangle = \langle n(\vec{k}_n,
\sigma_n)|\partial^{\mu}V^{(-)}_{\mu}(0) |p(\vec{k}_p,
\sigma_p)\rangle = 0
\end{eqnarray}
even for different masses of the neutron and proton. In the matrix
elements Eqs.(\ref{eq:A.3}) and (\ref{eq:A.4}) we have neglected the
contributions of the weak magnetism and one--pion exchange
\cite{Leitner2006}. In the approximation, when the squared momentum
transferred $q^2 = (\pm k_p \mp k_n)^2$ is much smaller than the
scales $M^2_V$ and $M^2_A$ defining the effective radii of the vector
and axial--vector form factors, the matrix elements Eqs.(\ref{eq:A.3})
and (\ref{eq:A.4}) can be reduced to the form
\begin{eqnarray}\label{eq:A.6}
\hspace{-0.3in}\langle p(\vec{k}_p, \sigma_p)|J^{(+)}_{\mu}(0)
|n(\vec{k}_n, \sigma_n)\rangle = \bar{u}_p(\vec{k}_p,
\sigma_p)\Big(\gamma_{\mu}(1 + \lambda \gamma^5) -
\frac{q_{\mu}\slashed{q}}{q^2}\Big)\,u_n(\vec{k}_n, \sigma_n)
\end{eqnarray}
and 
\begin{eqnarray}\label{eq:A.7}
\hspace{-0.3in}\langle n(\vec{k}_n, \sigma_n)|J^{(-)}_{\mu}(0)
|p(\vec{k}_p, \sigma_p)\rangle = \bar{u}_n(\vec{k}_n,
\sigma_n)\Big(\gamma_{\mu}(1 + \lambda \gamma^5) -
\frac{q_{\mu}\slashed{q}}{q^2}\Big)\,u_p(\vec{k}_p, \sigma_p),
\end{eqnarray}
where $\lambda = -1.2750(9)$ is the axial coupling constant
\cite{Abele2008} (see also
\cite{Ivanov2013,Ivanov2014,Ivanov2013a,Ivanov2017e}).

\begin{figure}
\includegraphics[height=0.155\textheight]{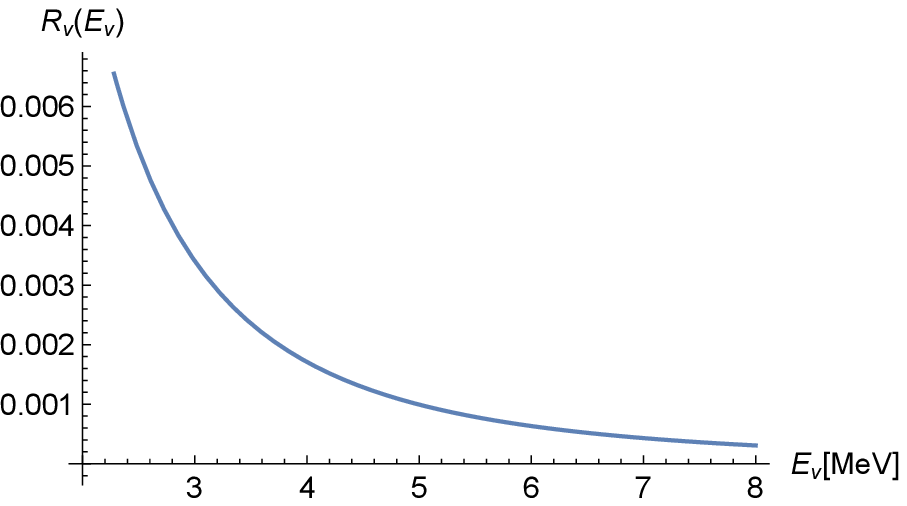}
\includegraphics[height=0.155\textheight]{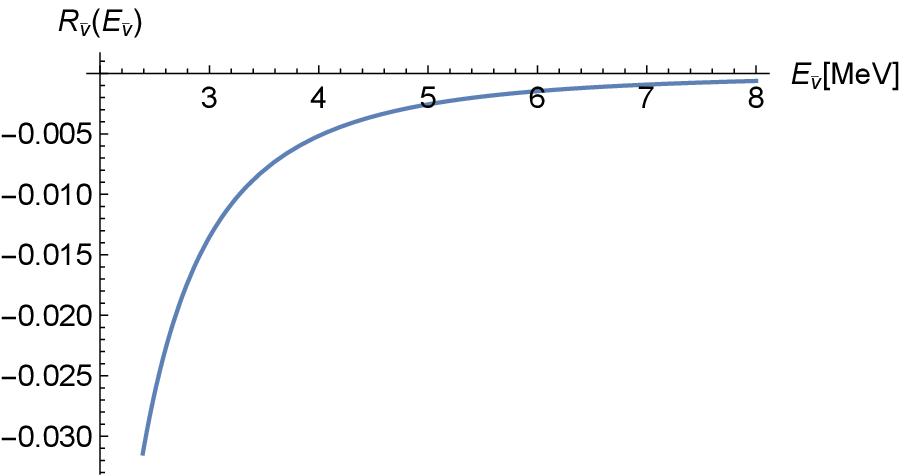}
  \caption{The relative contributions $R_{\nu}(E_{\nu})$ (left) and
    $R_{\bar{\nu}}(E_{\bar{\nu}})$ (right) of the ECVC effect to the
    cross sections for the inelastic electron neutrino--neutron
    and inverse $\beta$--decay in the neutrino and antineutrino energy
    regions $2\,{\rm MeV} \le E_{\nu} \le 8\,{\rm MeV}$ and $2\,{\rm
      MeV} \le E_{\bar{\nu}} \le 8\,{\rm MeV}$, calculated for
    $\lambda = - 1.2750$ \cite{Ivanov2013a} }
\label{fig:fig2}
\end{figure}

In order to illustrate the contribution of the term $-
q_{\mu}\slashed{q}/q^2$ responsible for the fulfilment of the
constraints Eq.(\ref{eq:A.5}), we neglect the contributions of the weak
magnetism, recoil and radiative corrections
\cite{Ivanov2013a}. Skipping intermediate standard calculations
\cite{Ivanov2013a} we obtain the following cross sections for the
inelastic electron neutrino--neutron scattering $\sigma(E_{\nu})$
and the inverse $\beta$--decay $\sigma(E_{\bar{\nu}})$:
\begin{eqnarray}\label{eq:A.8}
\sigma(E_{\nu}) &=& \sigma_0(E_{\nu}) +
\frac{G^2_F|V_{ud}|^2}{2\pi}\,k_- E_-\Big\{- \frac{m^2_e\Delta}{k_-
  E_- E_{\nu}}\Big[{\ell n}\Big(1 + \frac{2 k_-E_{\nu}\beta_- }{m^2_e
    - 2 E_- E_{\nu}}\Big) - {\ell n}\Big(1 - \frac{2 k_-E_{\nu}
  }{m^2_e - 2 E_- E_{\nu}}\Big)\Big]\nonumber\\ &+&
\frac{2m^2_e\Delta^2}{(m^2_e - 2 E_- E_{\nu})^2 - 4 k^2_- E^2_{\nu}} -
\frac{m^2_e \Delta^2}{4 k_- E_- E^2_{\nu}}\Big[{\ell n}\Big(1 +
  \frac{2 k_-E_{\nu}\beta_- }{m^2_e - 2 E_- E_{\nu}}\Big) - {\ell
    n}\Big(1 - \frac{2 k_-E_{\nu} }{m^2_e - 2 E_-
    E_{\nu}}\Big)\nonumber\\ &-&\frac{4 k_- E_{\nu}(m^2_e - 2 E_-
    E_{\nu}) }{(m^2_e - 2 E_- E_{\nu})^2 - 4 k^2_-
    E^2_{\nu}}\Big]\Big\},
\end{eqnarray}
where $E_- = E_{\nu} + \Delta$, $k_- = \sqrt{E^2_- - m^2_e}$ and
$\beta_- = k_-/E_-$ are the energy, momentum and velocity of the
electron, and
\begin{eqnarray}\label{eq:A.9}
\sigma(E_{\bar{\nu}}) &=& \sigma_0(E_{\bar{\nu}}) +
\frac{G^2_F|V_{ud}|^2}{2\pi}\,k_+ E_+\Big\{\frac{m^2_e\Delta}{k_+ E_+
  E_{\bar{\nu}}}\Big[{\ell n}\Big(1 + \frac{2 k_+ E_{\bar{\nu}}
  }{m^2_e - 2 E_+ E_{\bar{\nu}}}\Big) - {\ell n}\Big(1 - \frac{2
    k_+ E_{\bar{\nu}} }{m^2_e - 2 E_+
    E_{\bar{\nu}}}\Big)\Big]\nonumber\\ &+&
\frac{2m^2_e\Delta^2}{(m^2_e - 2 E_+ E_{\bar{\nu}})^2 - 4 k^2_+
  E^2_{\bar{\nu}}} - \frac{m^2_e \Delta^2}{4 k_+ E_+ E^2_{\bar{\nu}}}\Big[{\ell
    n}\Big(1 + \frac{2 k_+ E_{\bar{\nu}} }{m^2_e - 2 E_+ E_{\bar{\nu}}}\Big) -
  {\ell n}\Big(1 - \frac{2 k_+ E_{\bar{\nu}} }{m^2_e - 2 E_+
    E_{\bar{\nu}}}\Big)\nonumber\\ &-&\frac{4 k_+ E_{\bar{\nu}}(m^2_e - 2 E_+
    E_{\bar{\nu}}) }{(m^2_e - 2 E_+ E_{\bar{\nu}})^2 - 4 k^2_+
    E^2_{\bar{\nu}}}\Big]\Big\},
\end{eqnarray}
where $E_+ = E_{\bar{\nu}} - \Delta$ and $k_+ = \sqrt{E^2_+ - m^2_e}$
are the energy and momentum of the positron. The cross sections $
\sigma_0(E_{\nu})$ and $\sigma_0(E_{\bar{\nu}})$ are given by
\cite{Ivanov2013a}
\begin{eqnarray}\label{eq:A.10}
\sigma_0(E_{\nu}) = (1 + 3 \lambda^2)\,
\frac{G^2_F|V_{ud}|^2}{\pi}\,k_- E_-\quad,\quad
\sigma_0(E_{\bar{\nu}}) = (1 + 3 \lambda^2)\,
\frac{G^2_F|V_{ud}|^2}{\pi}\,k_+ E_+.
\end{eqnarray}
In the inelastic electron neutrino--neutron scattering and the
inverse $\beta$--decay the energies of neutrino and antineutrino vary
in the regions $E_{\nu} \ge 0$ and $E_{\bar{\nu}} \ge
(E_{\bar{\nu}})_{\rm thr} = ((m_n + m_e)^2 - m^2_p)/2m_p =
1.8061\,{\rm MeV}$ \cite{Ivanov2013a}.  The terms dependent on
$\Delta$ are caused by the ECVC effect. The relative contributions of
the ECVC effect to the cross sections under consideration we define as
follows
\begin{eqnarray}\label{eq:A.11}
\hspace{-0.3in}&&R_{\nu}(E_{\nu}) = \frac{1}{2}\,\frac{1}{1 +
  3\lambda^2}\Big\{- \frac{m^2_e\Delta}{k_- E_- E_{\nu}}\Big[{\ell
    n}\Big(1 + \frac{2 k_-E_{\nu}\beta_- }{m^2_e - 2 E_- E_{\nu}}\Big)
  - {\ell n}\Big(1 - \frac{2 k_-E_{\nu} }{m^2_e - 2 E_-
    E_{\nu}}\Big)\Big]+ \frac{2m^2_e\Delta^2}{(m^2_e - 2 E_-
  E_{\nu})^2 - 4 k^2_- E^2_{\nu}}\nonumber\\
\hspace{-0.3in}&& - \frac{m^2_e
  \Delta^2}{4 k_- E_- E^2_{\nu}}\Big[{\ell n}\Big(1 + \frac{2
    k_-E_{\nu}\beta_- }{m^2_e - 2 E_- E_{\nu}}\Big) - {\ell n}\Big(1 -
  \frac{2 k_-E_{\nu} }{m^2_e - 2 E_-
    E_{\nu}}\Big) -\frac{4 k_- E_{\nu}(m^2_e - 2 E_-
    E_{\nu}) }{(m^2_e - 2 E_- E_{\nu})^2 - 4 k^2_-
    E^2_{\nu}}\Big]\Big\},
\end{eqnarray}
and 
\begin{eqnarray}\label{eq:A.12}
\hspace{-0.3in}&&R_{\bar{\nu}}(E_{\bar{\nu}}) =
\frac{1}{2}\,\frac{1}{1 + 3\lambda^2} \Big\{\frac{m^2_e\Delta}{k_+ E_+
  E_{\bar{\nu}}}\Big[{\ell n}\Big(1 + \frac{2 k_+ E_{\bar{\nu}}
  }{m^2_e - 2 E_+ E_{\bar{\nu}}}\Big) - {\ell n}\Big(1 - \frac{2 k_+
    E_{\bar{\nu}} }{m^2_e - 2 E_+ E_{\bar{\nu}}}\Big)\Big] +
\frac{2m^2_e\Delta^2}{(m^2_e - 2 E_+ E_{\bar{\nu}})^2 - 4 k^2_+
  E^2_{\bar{\nu}}}\nonumber\\
\hspace{-0.3in}&& - \frac{m^2_e \Delta^2}{4 k_+ E_+
  E^2_{\bar{\nu}}}\Big[{\ell n}\Big(1 + \frac{2 k_+ E_{\bar{\nu}}
  }{m^2_e - 2 E_+ E_{\bar{\nu}}}\Big) - {\ell n}\Big(1 - \frac{2 k_+
    E_{\bar{\nu}} }{m^2_e - 2 E_+ E_{\bar{\nu}}}\Big) - \frac{4 k_+
    E_{\bar{\nu}}(m^2_e - 2 E_+ E_{\bar{\nu}}) }{(m^2_e - 2 E_+
    E_{\bar{\nu}})^2 - 4 k^2_+ E^2_{\bar{\nu}}}\Big]\Big\},
\end{eqnarray}
where $R_{\nu}(E_{\nu}) = \Delta \sigma(E_{\nu})/\sigma_0(E_{\nu})$,
$R_{\bar{\nu}}(E_{\bar{\nu}}) = \Delta
\sigma(E_{\bar{\nu}})/\sigma_0(E_{\bar{\nu}})$ with $\Delta
\sigma(E_{\nu}) = \sigma(E_{\nu}) - \sigma_0(E_{\nu})$ and $\Delta
\sigma(E_{\bar{\nu}}) = \sigma(E_{\bar{\nu}}) -
\sigma_0(E_{\bar{\nu}})$, respectively. The cross sections
Eq.(\ref{eq:A.8}) and Eq.(\ref{eq:A.9}) are calculated in the
laboratory frame in the non--relativistic approximation for outgoing
hadrons. Since the most important region of the antineutrino energies
for the inverse $\beta$--decay is $2\,{\rm MeV} \le E_{\bar{\nu}} \le
8\,{\rm MeV}$ \cite{Ivanov2013a}, in Fig.\,\ref{fig:fig2} we plot
$R_{\nu}(E_{\nu})$ and $R_{\bar{\nu}}(E_{\bar{\nu}})$ for $E_{\nu}$
and $E_{\bar{\nu}}$ varying over the regions $2\,{\rm MeV} \le E_{\nu}
\le 8\,{\rm MeV}$ and $2\,{\rm MeV} \le E_{\bar{\nu}} \le 8\,{\rm
  MeV}$, respectively.

Our numerical analysis of the relative contributions of the ECVC
effect to the cross sections for the inelastic electron
neutrino--neutron scattering and for the inverse $\beta$--decay shows
that these processes are not sensitive to the ECVC effect.  Indeed,
the contribution of the ECVC effect to the cross section for the
inelastic electron neutrino--neutron scattering is smaller than
$0.7\,\%$ at $E_{\nu} \simeq 2\,{\rm MeV}$ and decreases by about two
orders of magnitude at $E_{\nu} \simeq 8\,{\rm MeV}$. The cross
section for the inverse $\beta$--decay, applied to the analysis of the
deficit of positrons induced by reactor electron antineutrinos
\cite{Ivanov2013a,Mention2013}, should be averaged over the reactor
electron antineutrino energy spectrum, which has a maximum at
$E_{\bar{\nu}} \simeq 4\,{\rm MeV}$. According to
Fig.\,\ref{fig:fig2}, the contribution of the ECVC effect should
decease the yield of positrons $Y_{e^+}$ by about $0.5\,\%$. Since
such a contribution is smaller than the experimental error bars
$Y_{e^+} = 0.943(23)$ \cite{Mention2013}, one may argue that the
inverse $\beta$--decay is insensitive to the contribution of the ECVC
effect.

\section{Appendix B: The Lorentz structure of the  matrix element of
 the hadronic $n \to p$ transition, caused by the charged hadronic
 axial--vector current}
\renewcommand{\theequation}{B-\arabic{equation}}
\setcounter{equation}{0}

In this Appendix we analyse the Lorentz structure of the axial--vector
part of the hadronic $n \to p$ transition, induced by the charged
hadronic axial--vector current
\begin{eqnarray}\label{eq:B.1}
A^{(+)}_{\mu}(x) = \bar{N}(x)\tau^{(+)}\gamma_{\mu}\gamma^5 N(x).
\end{eqnarray}
The matrix element of our interest is 
\begin{eqnarray}\label{eq:B.2}
\langle {\rm out},
p(\vec{k}_p,\sigma_p)|A^{(+)}_{\mu}(0)|{\rm in}, n(\vec{k}_n,
\sigma_n)\rangle,
\end{eqnarray}
where $\langle {\rm out}, p(\vec{k}_p,\sigma_p)|$ and $|{\rm in},
n(\vec{k}_n, \sigma_n)\rangle$ are the wave functions of the free
proton and neutron in the final (i.e. out--state at $t \to + \infty$)
and initial (i.e. in--state at $t \to - \infty$) states, respectively
\cite{Itzykson1980}. Using the relation $\langle {\rm out},
p(\vec{k}_p,\sigma_p)| = \langle {\rm in},
p(\vec{k}_p,\sigma_p)|{\mathbb S}$, where ${\mathbb S}$ is the
S--matrix, we rewrite the matrix element Eq.(\ref{eq:B.2}) as follows
\begin{eqnarray}\label{eq:B.3}
\langle {\rm out}, p(\vec{k}_p,\sigma_p)|A^{(+)}_{\mu}(0)|{\rm in},
n(\vec{k}_n, \sigma_n)\rangle = \langle {\rm in},
p(\vec{k}_p,\sigma_p)|{\mathbb S} A^{(+)}_{\mu}(0)|{\rm in},
  n(\vec{k}_n, \sigma_n)\rangle.
\end{eqnarray}
Since the transition $n \to p$ is fully induced by strong low--energy
interactions, we define the S--matrix only in
terms of strong low--energy interaction described by the Lagrangian
Eq.(\ref{eq:4}). It is given by (see Eq.(\ref{eq:12}). Plugging Eq.(\ref{eq:12}) into Eq.(\ref{eq:B.3}) we get 
\begin{eqnarray}\label{eq:B.4}
\langle {\rm in}, p(\vec{k}_p,\sigma_p)|{\mathbb S}
A^{(+)}_{\mu}(0)|{\rm in}, n(\vec{k}_n, \sigma_n)\rangle = \langle
{\rm in}, p(\vec{k}_p,\sigma_p)|{\rm T}\Big(e^{\textstyle i\int
  d^4x\,{\cal L}_{\pi NN}(x)}A^{(+)}_{\mu}(0)\Big)|{\rm in},
n(\vec{k}_n, \sigma_n)\rangle.
\end{eqnarray}
After the integration over the pion--fields we arrive at the
expression
\begin{eqnarray}\label{eq:B.5}
\hspace{-0.3in}&&\langle {\rm out},
p(\vec{k}_p,\sigma_p)|A^{(+)}_{\mu}(0)|{\rm in}, n(\vec{k}_n,
\sigma_n)\rangle = {_N}\langle {\rm in}, p(\vec{k}_p,\sigma_p)|{\rm
  T}\Big\{\bar{N}(0)\tau^{(+)}\gamma_{\mu}\gamma^5
N(0)\,\exp\Big(i\,\frac{1}{2}\,g^2_{\pi}\int d^4x'
d^4y'\,\bar{N}(x')i\gamma^5 \vec{\tau} N(x') \cdot \nonumber\\
\hspace{-0.3in}&&\Delta(x' - y')\,\bar{N}(y')i\gamma^5 \vec{\tau}
N(y')\Big\}|{\rm in}, n(\vec{k}_n, \sigma_n)\Big)\rangle_N.
\end{eqnarray}
To order $g^2_{\pi}$ the dynamical contribution of the nucleon part of
the charge hadronic axial--vector current to the matrix element of the
transition $n \to p$, given by Eq.(\ref{eq:B.5}), is determined by the
matrix element
\begin{eqnarray}\label{eq:B.6}
\hspace{-0.3in}&&{_N}\langle {\rm in}, p(\vec{k}_p,\sigma_p)|{\rm
  T}\Big(\bar{N}(0)\tau^{(+)}\gamma_{\mu}\gamma^5 N(0)
i\,\frac{1}{2}\,g^2_{\pi}\int d^4x d^4y\,\bar{N}(x)i\gamma^5
\vec{\tau} N(x) \cdot \Delta(x - y)\,\bar{N}(y)i\gamma^5 \vec{\tau}
N(y)\Big)|{\rm in}, n(\vec{k}_n, \sigma_n)\rangle_N =\nonumber\\
\hspace{-0.3in}&& = [\bar{u}_p(\vec{k}_p, \sigma_p)\gamma_{\mu}\gamma^5
u_n(\vec{k}_n, \sigma_n)]\,3\,(-i)\, g^2_{\pi}\int d^4x d^4y\,{\rm
  tr}\{(-i)S_F(y - x) i\gamma^5 (-i)S_F(x- y) i\gamma^5\}\,\Delta(x -
y)\nonumber\\
\hspace{-0.3in}&& + [\bar{u}_p(\vec{k}_p, \sigma_p)i \gamma^5
  u_n(\vec{k}_n, \sigma_n)]\,(-i)\, g^2_{\pi}\int d^4x d^4y\,{\rm
  tr}\{\gamma_{\mu} \gamma^5 (-i)S_F(- x) i\gamma^5 (-i)S_F(x)
\}\,\Delta(x - y)\,e^{\, i(k_p - k_n)\cdot y}\nonumber\\
\hspace{-0.3in}&& + [\bar{u}_p(\vec{k}_p, \sigma_p)i \gamma^5
  u_n(\vec{k}_n, \sigma_n)]\,(-i)\, g^2_{\pi}\int d^4x d^4y\,{\rm
  tr}\{\gamma_{\mu} \gamma^5 (-i)S_F(- y) i\gamma^5 (-i)S_F(y)
\}\,\Delta(x - y)\,e^{\, i(k_p - k_n)\cdot x}.
\end{eqnarray}
In the momentum representation we get
\begin{eqnarray}\label{eq:B.7}
\hspace{-0.3in}&&{_N}\langle {\rm in}, p(\vec{k}_p,\sigma_p)|{\rm
  T}\Big(\bar{N}(0)\tau^{(+)}\gamma_{\mu}\gamma^5 N(0)
i\,\frac{1}{2}\,g^2_{\pi}\int d^4x d^4y\,\bar{N}(x)i\gamma^5
\vec{\tau} N(x) \cdot \Delta(x - y)\,\bar{N}(y)i\gamma^5 \vec{\tau}
N(y)\Big)|{\rm in}, n(\vec{k}_n, \sigma_n)\rangle_N =\nonumber\\
\hspace{-0.3in}&& = [\bar{u}_p(\vec{k}_p,
  \sigma_p)\gamma_{\mu}\gamma^5 u_n(\vec{k}_n,
  \sigma_n)]\,3\,g^2_{\pi}\,\delta^{(4)}(0)\int \frac{d^4p
  d^4Q}{(2\pi)^4i}\,{\rm tr}\Big\{\frac{1}{m_N - \slashed{p} -
  i0}\,\gamma^5 \frac{1}{m_N - \slashed{p}- \slashed{Q} -
  i0}\,\gamma^5\Big\}\,\frac{1}{m^2_{\pi} - Q^2 - i0}\nonumber\\
\hspace{-0.3in}&& + [\bar{u}_p(\vec{k}_p, \sigma_p)\gamma^5
  u_n(\vec{k}_n, \sigma_n)]\, \frac{2 g^2_{\pi}}{m^2_{\pi} - q^2 - i0}
\int \frac{d^4p}{(2\pi)^4i}\,{\rm tr}\Big\{\gamma_{\mu}\gamma^5
\frac{1}{m_N - \slashed{p} - i0}\,\gamma^5\,\frac{1}{m_N - \slashed{p}
  - \slashed{q} - i0}\Big\},
\end{eqnarray}
where $q = k_p - k_n$. The contribution proportional to
$\delta^{(4)}(0)$ can be in principle removed by using the
normal--ordered form of the four--nucleon operator of interaction
\cite{Itzykson1980}. Indeed, replacing $\bar{N}(x)i\gamma^5 \vec{\tau}
N(x) \cdot \Delta(x - y)\,\bar{N}(y)i\gamma^5 \vec{\tau} N(y)$ by
$:\bar{N}(x)i\gamma^5 \vec{\tau} N(x) \cdot \Delta(x -
y)\,\bar{N}(y)i\gamma^5 \vec{\tau} N(y):$ the vacuum expectation value
of the operator $:\bar{N}(x)i\gamma^5 \vec{\tau} N(x) \cdot \Delta(x -
y)\,\bar{N}(y)i\gamma^5 \vec{\tau} N(y):$ is equal to zero. The
contribution of the momentum integral of the second term is divergent
and proportional to $q_{\mu}$. As a result, the matrix element
Eq.(\ref{eq:B.7} we may define as follows
\begin{eqnarray}\label{eq:B.8}
\hspace{-0.3in}&&{_N}\langle {\rm in}, p(\vec{k}_p,\sigma_p)|{\rm
  T}\Big(\bar{N}(0)\tau^{(+)}\gamma_{\mu}\gamma^5 N(0)
i\,\frac{1}{2}\,g^2_{\pi}\int d^4x d^4y\,\bar{N}(x)i\gamma^5
\vec{\tau} N(x) \cdot \Delta(x - y)\,\bar{N}(y)i\gamma^5 \vec{\tau}
N(y)\Big)|{\rm in}, n(\vec{k}_n, \sigma_n)\rangle_N =\nonumber\\
\hspace{-0.3in}&& =
\frac{g^2_{\pi}}{4\pi^2}\,D_N\,[\bar{u}_p(\vec{k}_p,
  \sigma_p)\gamma_{\mu}\gamma^5 u_n(\vec{k}_n, \sigma_n)] +
\frac{g^2_{\pi}}{4\pi^2}\,\frac{2 m_N q_{\mu}}{m^2_{\pi} - q^2 -
  i0}\,E_N\,[\bar{u}_p(\vec{k}_p, \sigma_p)\gamma^5 u_n(\vec{k}_n,
  \sigma_n)].
\end{eqnarray}
Thus, for the matrix element Eq.(\ref{eq:B.5}), calculated to order
$g^2_{\pi}$, we obtain the following expression
\begin{eqnarray}\label{eq:B.9}
\hspace{-0.3in}&&\langle {\rm out},
p(\vec{k}_p,\sigma_p)|A^{(+)}_{\mu}(0)|{\rm in}, n(\vec{k}_n,
\sigma_n)\rangle = {_N}\langle {\rm in}, p(\vec{k}_p,\sigma_p)|{\rm
  T}\Big\{\bar{N}(0)\tau^{(+)}\gamma_{\mu}\gamma^5
N(0)\,\exp\Big(i\,\frac{1}{2}\,g^2_{\pi}\int d^4x'
d^4y'\,\bar{N}(x')i\gamma^5 \vec{\tau} N(x') \cdot \nonumber\\
\hspace{-0.3in}&&\Delta(x' - y')\,\bar{N}(y')i\gamma^5 \vec{\tau}
N(y')\Big\}|{\rm in}, n(\vec{k}_n, \sigma_n)\Big)\rangle_N =\nonumber\\
\hspace{-0.3in}&&= \bar{u}_p(\vec{k}_p, \sigma_p)\Big(\Big(1 +
\frac{g^2_{\pi}}{4\pi^2}\,D_N\Big)\gamma_{\mu}\gamma^5 +
\frac{g^2_{\pi}}{4\pi^2}\,\frac{2 m_N q_{\mu}}{m^2_{\pi} - q^2 -
  i0}\,E_N\,\gamma^5\Big)\,u_n(\vec{k}_n, \sigma_n).
\end{eqnarray}
Following \cite{Landau1959} we may argue that the Lorentz structure of
the matrix element Eq.(\ref{eq:B.5}), calculated to order $g^2_{\pi}$,
should be valid to all order of the $g^2_{\pi}$--expansion. The latter
is, of course, because of invariance of strong low--energy
interactions under the $G$--transformations. Thus, the matrix element
Eq.(\ref{eq:B.5}) should have the following Lorentz structure, induced
by the first class axial--vector current \cite{Weinberg1958}
\begin{eqnarray}\label{eq:B.10}
\hspace{-0.3in}&&\langle {\rm out},
p(\vec{k}_p,\sigma_p)|A^{(+)}_{\mu}(0)|{\rm in}, n(\vec{k}_n,
\sigma_n)\rangle = \bar{u}_p(\vec{k}_p, \sigma_p)\Big( -
F_A(q^2)\,\gamma_{\mu}\gamma^5 + \frac{2 m_N q_{\mu}}{m^2_{\pi} - q^2
  - i0}\,F_P(q^2)\gamma^5\Big)\,u_n(\vec{k}_n, \sigma_n),
\end{eqnarray}
where $F_A(q^2)$ and $F_P(q^2)$ are the axial--vector and pseudoscalar
form factors \cite{Leitner2006}. Taking into account the PCAC
hypothesis (or the hypothesis of Partial Conservation of
Axial--vector Current) \cite{Adler1968,DeAlfaro1973} we may rewrite the
r.h.s. of Eq.(\ref{eq:B.10}) as follows
\begin{eqnarray}\label{eq:B.11}
\hspace{-0.3in}&&\langle {\rm out},
p(\vec{k}_p,\sigma_p)|A^{(+)}_{\mu}(0)|{\rm in}, n(\vec{k}_n,
\sigma_n)\rangle = - \bar{u}_p(\vec{k}_p,
\sigma_p)\Big(\gamma_{\mu}\gamma^5 + \frac{2 m_N q_{\mu}}{m^2_{\pi} -
  q^2 - i0}\,\gamma^5\Big)\,F_A(q^2)\,u_n(\vec{k}_n, \sigma_n),
\end{eqnarray}
where we have set $F_A(q^2)= - F_P(q^2)$. At $q^2 = 0$ we set
$F_A(q^2) = \lambda$, where $\lambda = -1.2750(9)$ is the axial
coupling constant \cite{Abele2008} (see also
\cite{Ivanov2013,Ivanov2014,Ivanov2013a,Ivanov2017e}). In the chiral
limit $m_{\pi} \to 0$ the matrix element Eq.(\ref{eq:B.11}) multiplied
by the 4--momentum transferred $q^{\mu}$ vanishes
\begin{eqnarray}\label{eq:B.12}
\hspace{-0.3in}\lim_{m_{\pi} \to 0}q^{\mu}\langle {\rm out},
p(\vec{k}_p,\sigma_p)|A^{(+)}_{\mu}(0)|{\rm in}, n(\vec{k}_n,
\sigma_n)\rangle = 0
\end{eqnarray}
even for different masses of the neutron and proton, according to the
PCAC hypothesis \cite{Adler1968,DeAlfaro1973} pointing out an operator
relation $\partial^{\mu}\vec{A}_{\mu}(x) = m^2_{\mu}
F_{\pi}\,\vec{\pi}(x)$, where $F_{\pi}$ is the pion--decay constant
\cite{PDG2016}. In the chiral limit $m_{\pi} \to $ we get
$\partial^{\mu}\vec{A}_{\mu}(x) = 0$ agreeing well with
Eq.(\ref{eq:B.12}).

\end{document}